\documentclass[aps,prd,twocolumn,showpacs,superscriptaddress]{revtex4-1}  
\usepackage{setspace}
\usepackage{graphicx}  
\usepackage{dcolumn}   
\usepackage{bm}        
\usepackage{amssymb}   

\newcommand{\dzero}     {D0}
\newcommand{\ttbar}     {\mbox{$t\bar{t}$}}

\newcommand{\ppbar}     {\mbox{$p\bar{p}$}}
\newcommand{\bbbar}     {\mbox{$b\bar{b}$}}
\newcommand{\met}       {\mbox{$\not\!\!E_T$}}
\newcommand{\rar}       {\rightarrow}

\newcommand{\herwig}    {{\sc herwig}}
\newcommand{\pythia}    {{\sc pythia}}
\newcommand{\alpgen}    {{\sc alpgen}}

\newcommand{\mcnlo}	{{\sc mc@nlo}}
\newcommand{\geant}     {{\sc geant}}
\newcommand{\cteq}	{CTEQ6M}

\newcommand{\nuwt}	{$\nu$WT}
\newcommand{\mtfit}	{$m_t^{\rm fit}$}
\newcommand{\etal}	{\textit{et al.}}
\newcommand{\ljets}	{$\ell+$jets}
\newcommand{\gamjet}	{$\gamma+$jet}
\newcommand{\wjj}	{$W\rightarrow jj$}

\begin{document}


\hspace{5.2in} \mbox{Fermilab-Pub-12/020-E}

\title{Measurement of the Top Quark Mass in $\boldsymbol{\ppbar}$ Collisions using Events with Two Leptons}
\affiliation{Universidad de Buenos Aires, Buenos Aires, Argentina}
\affiliation{LAFEX, Centro Brasileiro de Pesquisas F{\'\i}sicas, Rio de Janeiro, Brazil}
\affiliation{Universidade do Estado do Rio de Janeiro, Rio de Janeiro, Brazil}
\affiliation{Universidade Federal do ABC, Santo Andr\'e, Brazil}
\affiliation{Instituto de F\'{\i}sica Te\'orica, Universidade Estadual Paulista, S\~ao Paulo, Brazil}
\affiliation{University of Science and Technology of China, Hefei, People's Republic of China}
\affiliation{Universidad de los Andes, Bogot\'{a}, Colombia}
\affiliation{Charles University, Faculty of Mathematics and Physics, Center for Particle Physics, Prague, Czech Republic}
\affiliation{Czech Technical University in Prague, Prague, Czech Republic}
\affiliation{Center for Particle Physics, Institute of Physics, Academy of Sciences of the Czech Republic, Prague, Czech Republic}
\affiliation{Universidad San Francisco de Quito, Quito, Ecuador}
\affiliation{LPC, Universit\'e Blaise Pascal, CNRS/IN2P3, Clermont, France}
\affiliation{LPSC, Universit\'e Joseph Fourier Grenoble 1, CNRS/IN2P3, Institut National Polytechnique de Grenoble, Grenoble, France}
\affiliation{CPPM, Aix-Marseille Universit\'e, CNRS/IN2P3, Marseille, France}
\affiliation{LAL, Universit\'e Paris-Sud, CNRS/IN2P3, Orsay, France}
\affiliation{LPNHE, Universit\'es Paris VI and VII, CNRS/IN2P3, Paris, France}
\affiliation{CEA, Irfu, SPP, Saclay, France}
\affiliation{IPHC, Universit\'e de Strasbourg, CNRS/IN2P3, Strasbourg, France}
\affiliation{IPNL, Universit\'e Lyon 1, CNRS/IN2P3, Villeurbanne, France and Universit\'e de Lyon, Lyon, France}
\affiliation{III. Physikalisches Institut A, RWTH Aachen University, Aachen, Germany}
\affiliation{Physikalisches Institut, Universit{\"a}t Freiburg, Freiburg, Germany}
\affiliation{II. Physikalisches Institut, Georg-August-Universit{\"a}t G\"ottingen, G\"ottingen, Germany}
\affiliation{Institut f{\"u}r Physik, Universit{\"a}t Mainz, Mainz, Germany}
\affiliation{Ludwig-Maximilians-Universit{\"a}t M{\"u}nchen, M{\"u}nchen, Germany}
\affiliation{Fachbereich Physik, Bergische Universit{\"a}t Wuppertal, Wuppertal, Germany}
\affiliation{Panjab University, Chandigarh, India}
\affiliation{Delhi University, Delhi, India}
\affiliation{Tata Institute of Fundamental Research, Mumbai, India}
\affiliation{University College Dublin, Dublin, Ireland}
\affiliation{Korea Detector Laboratory, Korea University, Seoul, Korea}
\affiliation{CINVESTAV, Mexico City, Mexico}
\affiliation{Nikhef, Science Park, Amsterdam, the Netherlands}
\affiliation{Radboud University Nijmegen, Nijmegen, the Netherlands and Nikhef, Science Park, Amsterdam, the Netherlands}
\affiliation{Joint Institute for Nuclear Research, Dubna, Russia}
\affiliation{Institute for Theoretical and Experimental Physics, Moscow, Russia}
\affiliation{Moscow State University, Moscow, Russia}
\affiliation{Institute for High Energy Physics, Protvino, Russia}
\affiliation{Petersburg Nuclear Physics Institute, St. Petersburg, Russia}
\affiliation{Instituci\'{o} Catalana de Recerca i Estudis Avan\c{c}ats (ICREA) and Institut de F\'{i}sica d'Altes Energies (IFAE), Barcelona, Spain}
\affiliation{Stockholm University, Stockholm and Uppsala University, Uppsala, Sweden}
\affiliation{Lancaster University, Lancaster LA1 4YB, United Kingdom}
\affiliation{Imperial College London, London SW7 2AZ, United Kingdom}
\affiliation{The University of Manchester, Manchester M13 9PL, United Kingdom}
\affiliation{University of Arizona, Tucson, Arizona 85721, USA}
\affiliation{University of California Riverside, Riverside, California 92521, USA}
\affiliation{Florida State University, Tallahassee, Florida 32306, USA}
\affiliation{Fermi National Accelerator Laboratory, Batavia, Illinois 60510, USA}
\affiliation{University of Illinois at Chicago, Chicago, Illinois 60607, USA}
\affiliation{Northern Illinois University, DeKalb, Illinois 60115, USA}
\affiliation{Northwestern University, Evanston, Illinois 60208, USA}
\affiliation{Indiana University, Bloomington, Indiana 47405, USA}
\affiliation{Purdue University Calumet, Hammond, Indiana 46323, USA}
\affiliation{University of Notre Dame, Notre Dame, Indiana 46556, USA}
\affiliation{Iowa State University, Ames, Iowa 50011, USA}
\affiliation{University of Kansas, Lawrence, Kansas 66045, USA}
\affiliation{Kansas State University, Manhattan, Kansas 66506, USA}
\affiliation{Louisiana Tech University, Ruston, Louisiana 71272, USA}
\affiliation{Boston University, Boston, Massachusetts 02215, USA}
\affiliation{Northeastern University, Boston, Massachusetts 02115, USA}
\affiliation{University of Michigan, Ann Arbor, Michigan 48109, USA}
\affiliation{Michigan State University, East Lansing, Michigan 48824, USA}
\affiliation{University of Mississippi, University, Mississippi 38677, USA}
\affiliation{University of Nebraska, Lincoln, Nebraska 68588, USA}
\affiliation{Rutgers University, Piscataway, New Jersey 08855, USA}
\affiliation{Princeton University, Princeton, New Jersey 08544, USA}
\affiliation{State University of New York, Buffalo, New York 14260, USA}
\affiliation{Columbia University, New York, New York 10027, USA}
\affiliation{University of Rochester, Rochester, New York 14627, USA}
\affiliation{State University of New York, Stony Brook, New York 11794, USA}
\affiliation{Brookhaven National Laboratory, Upton, New York 11973, USA}
\affiliation{Langston University, Langston, Oklahoma 73050, USA}
\affiliation{University of Oklahoma, Norman, Oklahoma 73019, USA}
\affiliation{Oklahoma State University, Stillwater, Oklahoma 74078, USA}
\affiliation{Brown University, Providence, Rhode Island 02912, USA}
\affiliation{University of Texas, Arlington, Texas 76019, USA}
\affiliation{Southern Methodist University, Dallas, Texas 75275, USA}
\affiliation{Rice University, Houston, Texas 77005, USA}
\affiliation{University of Virginia, Charlottesville, Virginia 22901, USA}
\affiliation{University of Washington, Seattle, Washington 98195, USA}
\author{V.M.~Abazov} \affiliation{Joint Institute for Nuclear Research, Dubna, Russia}
\author{B.~Abbott} \affiliation{University of Oklahoma, Norman, Oklahoma 73019, USA}
\author{B.S.~Acharya} \affiliation{Tata Institute of Fundamental Research, Mumbai, India}
\author{M.~Adams} \affiliation{University of Illinois at Chicago, Chicago, Illinois 60607, USA}
\author{T.~Adams} \affiliation{Florida State University, Tallahassee, Florida 32306, USA}
\author{G.D.~Alexeev} \affiliation{Joint Institute for Nuclear Research, Dubna, Russia}
\author{G.~Alkhazov} \affiliation{Petersburg Nuclear Physics Institute, St. Petersburg, Russia}
\author{A.~Alton$^{a}$} \affiliation{University of Michigan, Ann Arbor, Michigan 48109, USA}
\author{G.~Alverson} \affiliation{Northeastern University, Boston, Massachusetts 02115, USA}
\author{M.~Aoki} \affiliation{Fermi National Accelerator Laboratory, Batavia, Illinois 60510, USA}
% ...
%
%
\collaboration{The D0 Collaboration}\thanks{with visitors from
$^{a}$Augustana College, Sioux Falls, SD, USA,
$^{b}$The University of Liverpool, Liverpool, UK,
$^{c}$UPIITA-IPN, Mexico City, Mexico,
$^{d}$DESY, Hamburg, Germany,
$^{e}$SLAC, Menlo Park, CA, USA,
$^{f}$University College London, London, UK,
$^{g}$Centro de Investigacion en Computacion - IPN, Mexico City, Mexico,
$^{h}$ECFM, Universidad Autonoma de Sinaloa, Culiac\'an, Mexico,
and 
$^{i}$Universit{\"a}t Bern, Bern, Switzerland.
$^{\ddag}$Deceased.
} \noaffiliation
\vskip 0.25cm
\date{January 24, 2012}

\begin{abstract}
We present a measurement of the top-quark mass ($m_t$)
in \ppbar\ collisions at $\sqrt{s} = 1.96$ TeV using \ttbar\ events with 
two leptons ($ee$, $e\mu$, or $\mu\mu$) and accompanying jets
in $4.3$~fb$^{-1}$ of data collected with the \dzero\ 
detector at the Fermilab Tevatron collider.  
We analyze the kinematically underconstrained dilepton events
by integrating over their neutrino rapidity distributions. We reduce the dominant systematic uncertainties from the calibration of jet energy using a correction obtained from \ttbar\ events with a final state of a single lepton plus jets. We also correct jets in simulated events to replicate the quark flavor dependence of the jet response in data.  We measure $m_t = 173.7 \pm 2.8\thinspace(\rm stat) \pm 1.5\thinspace(\rm syst)$ GeV and combining with our analysis in 1 fb$^{-1}$ of preceding data we measure $m_t = 174.0 \pm 2.4\thinspace(\rm stat) \pm 1.4\thinspace(\rm syst)$ GeV.   Taking into account statistical and systematic correlations, a combination with the \dzero\ matrix element result from both data sets yields $m_t = 173.9 \pm 1.9\thinspace(\rm stat) \pm 1.6\thinspace(\rm syst)$ GeV.
\end{abstract}

\maketitle
\vskip 1.0in

The masses of fundamental fermions in the standard model
(SM) are generated through their interaction with a hypothesized scalar
Higgs field with a strength given by a Yukawa coupling specific 
to each fermion species.  
The Yukawa coupling of the top quark corresponds to unity 
within uncertainties,
and this value is constrained by a measurement of
the top-quark mass ($m_t$). In direct searches at the LHC for the standard model Higgs boson,
both the CMS and ATLAS experiments observe
local excesses above the background expectations for a Higgs
boson mass ($m_H$) of approximately 125 GeV/$c^2$~\cite{cmsh,atlash}, decaying to diboson
final states. Combined results in searches from the CDF and \dzero\
experiments at the Tevatron show evidence for events above background
expectation in \bbbar\ final states~\cite{tevh}. It is therefore important to
sharpen the measurement of $m_t$, as its precise value, along with the mass
of the $W$ boson ($m_W$), constrain the standard model prediction for $m_H$
through well defined radiative corrections.


In \ppbar\ collisions, top quarks ($t$) are primarily produced in
\ttbar\ pairs, 
with each top quark decaying with
$BR(t\rar Wb)\sim 100$\%. These events
yield final states with either 0, 1, or 2 leptons from 
decays of the two $W$ bosons.  We 
consider the dilepton channels ($2\ell$) that contain
either electrons or muons of large transverse momentum ($p_T$)
and at least two jets.
We analyzed such events previously~\cite{llmtop07,llmtop09}
using
the neutrino-weighting (\nuwt) approach~\cite{nuWTR1}. 
While the $2\ell$ channels have low background, the small decay 
branching ratio into leptons
means that $m_t$ measurements from these events remained 
statistically limited unlike in channels with one lepton and four or 
more jets (\ljets).  This situation 
has changed recently (e.g., Ref.~\cite{llmtopME}).
Now, dominant systematic uncertainties 
from jet energy calibration, which have been larger~\cite{llmtop09} 
in the dilepton channel compared to \ljets, 
are limiting precision of the $m_t$ measurement.  
In \ljets\ events, two quarks originate from $W$ boson decay
and yield a dijet mass signature that permits a precise
calibration of jet energies for the measurement of $m_t$
in \ttbar\ events~\cite{nature}.  
While this calibration
has greatly improved measurements in
the \ljets\ channels, it has not been carried over to
the calibration in other analyses.
This is primarily due to differences in event
topologies that can affect the details of the jet energy scale.

We present a new measurement of $m_t$ using the 
\dzero\ detector
with $4.3$ fb$^{-1}$ of \ppbar\ collider data 
in the $ee$, $e\mu$, and $\mu\mu$ final states.  
We improve the jet energy calibration for the accompanying
jets using the energy
scale from \ljets\ events~\cite{ljetsME}.
Our approach differs from that of Ref.~\cite{cdfllmtop} in that we
do not use the \ljets\ scale as a constraint in 
a combined fit of \ljets\ and dilepton events.  
Instead, we use this constraint as a calibration,
and estimate the uncertainties of transferring
that calibration to the dilepton event topology.
This procedure demonstrates how the calibration obtained using the dijet
constraint from $m_W$ can be applied to different final states, and
has wide applicability beyond the measurement of $m_t$ in $2\ell$ events.
We also employ flavor-dependent corrections
to jet energies for the first time in a dilepton analysis
that substantially reduce the uncertainties on jet energy
resulting from jet flavor.
The presented $m_t$ measurement is performed using the same
data as Ref.~\cite{llmtopME}, and is correlated with it
as discussed below.

The \dzero\ detector~\cite{d0nim} 
is a multipurpose detector operated 
at the Fermilab Tevatron \ppbar\ collider.  The inner detector 
consists of coaxial cylinders and disks
of silicon microstrips for track and vertex reconstruction.  Eight 
layers of scintillating fibers arranged in doublets surround the 
silicon microstrip tracker and extend tracking 
measurements
to forward pseudorapidities, $\eta$~\cite{pseudorapidity}.  
A 1.9 T solenoid produces 
a magnetic field for the tracking detectors.  Uranium-liquid argon 
calorimeters 
surround the tracking volume and perform both electromagnetic and 
hadronic shower energy measurements.  Thin scintillation intercryostat
detectors sample showers in the region between the central
and end calorimeters.  Three layers of proportional drift 
tubes and scintillation counters 
reside outside the calorimetry, with 1.8 T toroids that provide 
muon identification and independent measurement of muon momenta.

We simulate \ttbar\ events using Monte Carlo (MC) samples for 140 GeV
$<m_t< 200$~GeV using the \alpgen\ 
generator~\cite{alpgen} and 
\pythia~\cite{pythia} for parton fragmentation.  
Backgrounds originate from $Z/\gamma^*\rar 2\ell+$jets 
and $WW/WZ/ZZ\rar 2\ell+$jets production.  
For the former, we use \alpgen\ combined with \pythia, while diboson 
backgrounds
are simulated entirely with \pythia.  We pass all MC events through 
a full detector simulation based on \geant~\cite{geant}.
Backgrounds from instrumental effects that result 
in misidentified leptons are modeled using data.

We use single and two-lepton triggers to select events for
this analysis.  Data and simulated events are reconstructed to 
provide the momenta of tracks, jets, and
lepton candidates.  Charged leptons are required to be
isolated from other calorimeter energy deposits, 
and to have an associated track in the inner detector.
Calorimeter and tracking information are combined
to identify electrons.  Track 
parameters in the muon and inner detector system are used
to identify
muons.  We reconstruct jets with an iterative, midpoint cone 
algorithm with radius ${\cal R}_{\rm cone} = 0.5$~\cite{jetalgo}.
Jets are calibrated with the standard \dzero\
jet energy correction which is derived from data~\cite{jetcal}.
The method corrects the measured jet energy to the value
obtained by applying the reconstruction cone algorithm to
particles from jet fragmentation before they interact with the
detector.  We establish the efficacy of the method in the MC, 
where we compare the measured jet
and the jet reconstructed from fragmentation particles.
The jets in data and MC are calibrated independently so
that their relative response is close to unity.  This corrects for
detector response, energy deposited outside of the jet
cone, electronics noise, and pileup.  
The largest correction compensates for the detector response, 
and is extracted using \gamjet\ events in data and MC.  
We also correct jets for the $p_T$ of any embedded muon and that
of the associated neutrino.
We initially apply this standard calibration~\cite{jetcal} because it provides
detailed $p_T$ and $\eta$ dependent corrections.
It also provides distinct corrections to jets and 
the imbalance in event transverse momentum (\met)
because several components
(e.g., noise and out-of-cone effects) result from the jet reconstruction
algorithm rather than any undetected energy.  
In the $p_T$ range of jets found in \ttbar\ events, 
the uncertainty of the 
standard \dzero\ jet energy calibration averages 2\%, and
is dominated by systematics. 
Because the flavor dependence of jet energy calibration 
can yield one of the largest systematic uncertainties on our
measurement~\cite{llmtop09}, we have improved our analysis
by accounting for this dependence.
We use responses of single particles from data and MC to determine 
the energy scale for different jet flavors.  We correct MC jets 
by the ratio of data response
to MC response according to their flavor to ensure that the MC 
reflects the flavor dependence in data, as in Ref.~\cite{ljetsME}.
We calculate \met\ as
the negative of the vector sum of all transverse components of
calorimeter cell energies and muon track momenta, 
corrected for the response to electrons and jets.

Events are
selected to have two leptons ($ee$, $e\mu$, $\mu\mu$) and
two or more jets.  The leptons must have $p_T> 15$~GeV and
the jets must have $p_T>20$~GeV.  Electrons and jets are
required to satisfy $|\eta|<2.5$, while muons must have
$|\eta|<2$.  We further require
\met $>40$~GeV in the $\mu\mu$ channel.  The $e\mu$ events must satisfy
$H_T>120$ GeV, where $H_T$ is defined to be the sum of the
$p_T$s of jets and the leading lepton.  
In $\mu\mu$ and $ee$ events, we also require \met\ to be
significantly larger than typical values found in the
distribution from $Z$ boson events.
These and all other selections are detailed in Ref.~\cite{llcsec}.  
We observe 50, 198, and 84 events with expected background
yields of 10.4, 28.1, and 31.0 events in the $ee$, 
$e\mu$, and $\mu\mu$ channels, respectively.

In \ljets\ events, one $W$ boson decays to two
quarks that fragment to jets.  
The invariant mass of this jet pair can be used to improve
the calibration for all jets in these events.
Complications arise because 
the four jets in the \ljets\  
events can be incorrectly assigned to the initial four quarks.  Energy
from different partons is also mixed in the same jet due
to a high jet multiplicity.
Observed jet energies are also affected by color flow effects, 
which are different for the $b$-quark jets and for jets from the decay of
color singlet $W$ bosons.  These attributes 
are specific to a particular event topology such as 
\ljets.  Nevertheless, a scale factor based on the dijet 
invariant mass that is
correlated with $m_W$ can be extracted.
The most recent analysis of this kind by \dzero\ used 
2.6 fb$^{-1}$ of data and obtained a calibration
factor of $1.013\pm0.008\thinspace(\rm stat)$~\cite{ljetsME}.  
The uncertainty of 0.8\%
is smaller than that of the standard
jet energy correction and will decrease with additional
data. There are additional systematic effects on this energy scale that one must account for when 
applying it to $b$-quark jets in the \ljets\ analysis.  These also affect our analysis, and we
similarly evaluate the flavor dependence and residual energy scale systematic uncertainties directly on the measured $m_t$ to avoid double counting.  These are quoted in Table II and discussed below.  Beyond this, we have the possible difference between $b$-quark jets in dilepton events and $b$-quark jets in \ljets\ events and the effect of using a calibration based on a subset of the total data, each of which we discuss now in detail.

The event topology
is different in 
$2\ell$ and \ljets\ events.  
This has prevented significant progress in reducing the large
standard jet energy scale uncertainties in dilepton analyses.
To overcome this challenge and
carry over the \ljets\ calibration, we must 
account for the possibility that
the energy scale of the $b$-quark jets in the two channels can differ. We calculate the energy scale, $R^{2\ell}$, for $b$-quark jets in the dilepton sample using responses for single particles that fall within the reconstructed jet cone. This is done by scaling single particle responses in MC to reproduce the energy response of jets in data~\cite{jetcrosec}, giving $R_{\rm data}^{2\ell}$, and using particle responses from MC, giving $R_{\rm MC}^{2\ell}$.  We calculate the ratio of these two responses in the dilepton channel and the analogous ratio for $b$-quark jets in the \ljets\ sample. The corresponding double ratio
\begin{equation}
\label{eq:dratio}
{\cal R}^b_{2\ell}(p_T^b)
= \frac{R_{\rm \/data}^{2\ell}(p_T^b)/R_{\rm \/MC}^{2\ell}(p_T^b)}{R_{\rm \/data}^{\ell+{\rm jets}}(p_T^b)/R_{\rm \/
MC}^{\ell+{\rm jets}}(p_T^b)},
\end{equation}
varies between 1.001 and 1.003 depending
on $b$-quark jet $p_T$, $p_T^b$.  
The multiplicity of particles in $b$-quark jets in
\ljets\ events at the MC generator level is, after application
of the offline jet algorithm,
a few percent higher than in the dilepton sample,
which is a sufficiently large difference to account for the observed
value of ${\cal R}^b_{2\ell}$.  We therefore take 0.3\%, 
the maximum excursion of ${\cal R}^b_{2\ell}$ from unity,
as a systematic uncertainty on 
carrying over the \ljets\ scale to the jets in our 
dilepton sample.  The \ljets\ scale is applied as a direct correction 
to the standard calibration.

The jet energy scale 
calibration obtained in Ref.~\cite{ljetsME} is based on a subset 
of the data, and we must therefore estimate 
the effect of using the calibration on a larger data set.  The 
instantaneous luminosity of the dilepton sample is higher on
average.  We reweight the distribution of the number of primary 
vertices in the \ljets\ sample to match the distribution 
in the 4.3 fb$^{-1}$ \ljets\ data and recalculate
the \ljets\ energy scale.  This produces a negligible effect.  
To account for a possible shift in the energy scale of the liquid
argon calorimeter, we apply a correction derived from 4.3 
~fb$^{-1}$ rather than 2.6 fb$^{-1}$, and this 
yields a 0.7\% shift in jet energy scale.  
From these studies, we obtain a total uncertainty on
the \ljets\ energy scale as applied to our analysis as the sum in
quadrature of the statistical uncertainty (0.8\%), 
${\cal R}^b_{2\ell}$ (0.3\%), and the calorimeter calibration (0.7\%).
This yields a 1.1\% uncertainty for applying the \ljets\ energy scale.

The consequence of two neutrinos in dilepton events
is an underconstrained kinematics. We employ the \nuwt\
technique to extract $m_t$~\cite{nuWTR1} due to its weak sensitivity
to the modeling details of \ttbar\ events.  
We integrate over the $\eta$ distributions predicted for
both neutrinos, solve the event kinematics, and calculate 
\met\ from the neutrino momentum solutions. The expected neutrino $\eta$ distribution in the dilepton channel is symmetric around $\eta$=0 and found to be well-described by a Gaussian distribution.  The width of the distribution decreases gradually with increasing $m_t$ (i.e., as the neutrinos become more central). Hence, we model the neutrino $\eta$ distributions with a Gaussian probability distribution using a width parameterized as a linear function of $m_t$.
Several more sophisticated parametrizations were
tested, but provided negligible improvement in expected
precision in pseudoexperiments.
By comparing the calculated \met\ to the measured \met\
for each event, 
we calculate a weight for a given choice of $m_t$.  
For each neutrino rapidity sampling, we sum the weights 
calculated from all combinations of neutrino momentum
solutions and jet assignments.  We therefore
arrive at a distribution of relative weight for a range of 
$m_t$ for each event.  
We found in Ref.~\cite{llmtop09} that most
of the statistical sensitivity to $m_t$ is obtained from
the first two moments of this weight distribution, the mean
($\mu_w$) and RMS ($\sigma_w$).  A coarse granularity
of our sampling of the $\eta$ distribution causes these
moments to be unstable.  To reduce this variation, we 
have increased the sampling for
this integration by an order of magnitude relative to 
our previous analysis~\cite{llmtop09}.  This 
improves the expected
statistical uncertainty on $m_t$ by~4\%.
Requiring the integral of this distribution to be nonzero
excludes events with a measured \met\ that is incompatible
with coming from neutrinos from \ttbar\ decay.  This introduces
a small inefficiency for the \ttbar\ signal and reduces the 
background contamination in the final sample.
Our final kinematically reconstructed
data sample consists of 49, 190, and 80 events in the $ee$, $e\mu$, 
and $\mu\mu$ channels, respectively.

\begin{figure}
\includegraphics[scale=0.4]{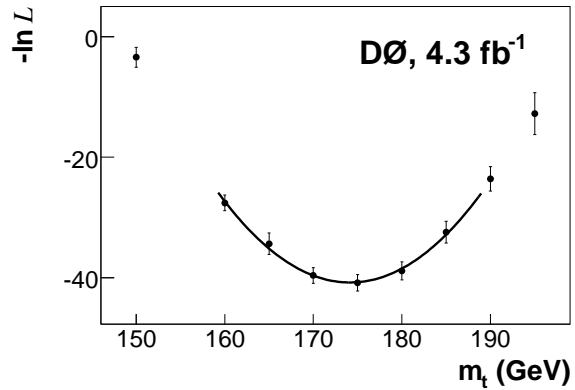}
\caption{\label{fig:like} 
-$\ln L$ as a function of $m_t$ for the combined
$ee$, $e\mu$, and $\mu\mu$ channels.  
A parabolic fit is shown near the minimum value in $m_t$.}
\end{figure}

Probability distributions for $\mu_w$ and $\sigma_w$
are constructed for 
background in each channel. Each background component is 
normalized to its expected event yield.  
We generate distributions of \ttbar\ signal probability as a
function of $\mu_w$, $\sigma_w$, and $m_t$. 
We use a binning
that provides the minimum expected statistical uncertainty, as checked
in pseudoexperiments.
We perform a binned maximum likelihood fit 
to the probability distributions, fixing the total 
signal and background yields expected in our data.  The signal 
is normalized to the cross section calculated for \ttbar\
production~\cite{mochUwer}, evaluated at $m_t = 172.5$~GeV.  
For all measurements, we obtain a likelihood
($L$) vs $m_t$.  We fit
a parabola to the dependence of $-\ln{L}$ vs. $m_t$,
and the fitted mass, \mtfit, is defined as the lowest point of the 
parabola.  Point-to-point fluctuations 
mean that the initial placement of the window may result in 
an nonconvergent fit.  
We therefore iterate the fit around
the current fit minimum.  This results in a significant
improvement in fitting efficiency, particularly in the
dimuon channel.
The final $-\ln{L}$ vs. $m_t$ for data is shown in 
Fig.~\ref{fig:like}.  The
statistical uncertainty for each measurement is taken as
the half-width of the parabola at 0.5 units in
$-\ln{L}$ above the minimum at \mtfit. 

The above procedure is followed for the extraction of $m_t$
from data and is used to calibrate the result as follows.
We construct pseudoexperiments from signal and background MC
samples according to their expected yields and allow
fluctuations in each such that the total 
equals the number of observed events.
We perform 1000 pseudoexperiments for each channel, and measure
\mtfit\ in each.  A linear fit of \mtfit\
vs the input $m_t$ provides a calibration for our method.
We also calculate the pull width of the average estimated statistical
uncertainty vs the rms of \mtfit\ values.
The resulting slopes, offsets, and pull widths are given
in Table~\ref{tab:calib}.
\begin{table}
\caption{\label{tab:calib} 
Parameters used to calibrate \mtfit\ in the analysis of $ee$,
$e\mu$, and $\mu\mu$ channels and their combination.}
\begin{ruledtabular}
\begin{tabular}{lccc}
Channel		& Slope 	& Offset [GeV]  & Pull width \\
\hline
$ee$		& $0.976\pm0.014$ & $0.03\pm0.16$ & $1.01\pm0.01$ \\
$e\mu$		& $0.973\pm0.012$ & $0.43\pm0.14$ & $1.03\pm0.01$ \\
$\mu\mu$	& $1.038\pm0.022$ & $0.49\pm0.23$ & $1.06\pm0.03$ \\
\end{tabular}
\end{ruledtabular}
\end{table}
The \mtfit\ and estimated statistical uncertainty
are corrected with these parameters.
We obtain a calibrated mass
measurement for the $4.3$ fb$^{-1}$ sample in the $ee$,
$e\mu$, and $\mu\mu$ channels.

The largest systematic uncertainties are associated with the
jet calibration.  We change
the \ljets\ energy scale factor by $\pm$1.1\%, and perform 
our analysis to obtain a systematic uncertainty on $m_t$
of 0.9 GeV.
The result of the \ljets\ analysis is a single scale
factor averaged over all jet $p_T$s that are utilized in the 
dijet mass, i.e. dominated by light quark jets from $W$ boson 
decay.  
As in Ref.~\cite{ljetsME}, we estimate an uncertainty due to
the difference in $p_T$ distributions of $b$-quark jets, in 
our case in dilepton events, vs the calibrating jets from the
\wjj\ sample.  To estimate an uncertainty from this difference, we 
treat the $p_T$ and $\eta$
dependence of the uncertainty in the standard jet energy scale 
as a possible dependence of the residual energy scale 
following the calibration to \ljets. 
We calculate the average of the energy scale uncertainty for
jets in the \wjj\ sample.  For each jet in the dilepton
sample, we apply a shift
corresponding to the difference between its uncertainty in 
energy scale and the \wjj\ sample's average uncertainty in 
energy scale.  Propagating this
difference through the mass analysis yields a 0.3 GeV
uncertainty on $m_t$.

The flavor-dependent jet energy corrections described earlier 
provide MC-based mass templates that
accurately reflect the data.  As in \cite{ljetsME},
we propagate the uncertainty in these corrections and obtain a systematic 
uncertainty on $m_t$ of 0.5 GeV. The uncertainties due to flavor dependence and residual scale together with the uncertainty originating from the carry over of the jet energy scale from the \ljets\ sample account for the difference between $b$-quark jets in dilepton events and jets from \wjj\ in \ljets\ events.

We evaluate the effect of our uncertainty in modeling 
initial state radiation (ISR) and final state radiation 
(FSR) by comparing two \pythia\ samples having identical values of generated
$m_t$ but different input parameters taken from a CDF study~\cite{tevtopcomb} corresponding to an increased
or decreased amount of ISR/FSR. 
Color reconnection uncertainties are estimated 
by comparing the analysis with \pythia\ Tune Apro and \pythia\ Tune ACpro using~\cite{tuneApro}.  
Higher order QCD evolution is estimated by 
comparing \alpgen\ configured with \pythia\ to \mcnlo\ with 
\herwig~\cite{herwig} and this accounts for the uncertainty due to underlying event as well. To estimate sensitivity to
uncertainties in the parton distribution functions, we
use \cteq, and employ the method described in 
Ref.~\cite{cteq}.

We modify the jet energy resolution in MC events to
reflect the resolution in data. We evaluate the effect of an 
uncertainty in this procedure on the extraction of $m_t$ by 
shifting the jet resolution by one standard deviation.
We treat the electron and 
muon energy and momentum scales similarly and shift their 
calibrations within their uncertainties.

Pseudoexperiments are used similarly to account for the 
uncertainty in the method that arises from the
uncertainties on the
offset and slope in the calibration of the fitted $m_t$.
We estimate the uncertainty due to the statistics 
employed in our templates of the \ttbar\ probability distributions.  
We construct 1000 new templates, for both signal and background, 
and vary their bin contents within their Gaussian uncertainties.  
With these templates, we obtain 1000 new measurements from data
and quote the rms of these values as a systematic 
uncertainty.  We assign a systematic uncertainty on the 
signal fraction by shifting the background contributions in
pseudoexperiments within their total uncertainty.

\begin{table}
\caption{\label{tab:syst} 
Estimated systematic uncertainties on $m_t$ 
for the combined dilepton measurement in 4.3~fb$^{-1}$.}
\begin{ruledtabular}
\begin{tabular}{lc}
Source 					& Uncertainty (GeV) \\
\hline
Jet energy calibration				&  \\
\hskip 0.5cm Overall scale           		& 0.9 \\
\hskip 0.5cm Flavor dependence			& 0.5 \\
\hskip 0.5cm Residual scale			& 0.3 \\
Signal modeling					&  \\
\hskip 0.5cm ISR/FSR				& 0.4 \\
\hskip 0.5cm Color reconnection		        & 0.5 \\
\hskip 0.5cm Higher order effects		& 0.6 \\
\hskip 0.5cm $b$ quark fragmentation		& 0.1 \\
\hskip 0.5cm PDF uncertainty			& 0.5 \\
Object reconstruction				&  \\
\hskip 0.5cm Muon $p_T$ resolution		& 0.2 \\
\hskip 0.5cm Electron energy scale		& 0.2 \\
\hskip 0.5cm Muon $p_T$ scale			& 0.2 \\
\hskip 0.5cm Jet resolution			& 0.3 \\
\hskip 0.5cm Jet identification		        & 0.3 \\
Method						&  \\
\hskip 0.5cm Calibration			& 0.1 \\
\hskip 0.5cm Template statistics		& 0.5 \\
\hskip 0.5cm Signal fraction			& 0.2 \\
\hline
Total systematic uncertainty			& 1.5 \\
\end{tabular}
\end{ruledtabular}
\end{table}

We combine measurements in the three dilepton channels using the
method of ``best linear unbiased estimator"~\cite{blue}.
We calculate each systematic uncertainty 
for the combined result, as given in Table~\ref{tab:syst}, 
according to its correlation among channels.  The
resulting measurement gives 
$m_t = 173.7\pm2.8\thinspace(\rm stat)\pm1.5\thinspace(\rm syst)$~GeV.

We combine this measurement with \dzero's measurement in the preceding
1 fb$^{-1}$ of data using the \nuwt\ and matrix weighting 
methods~\cite{llmtop09}.  Some uncertainties
evaluated in the 4.3 fb$^{-1}$ sample are not available for
the 1.0 fb$^{-1}$ sample. We include
the new uncertainties in
the result from the previous analysis.  We 
consider statistical uncertainties, as well as the following
systematic uncertainties to be uncorrelated: 
calibration of method, template statistics,
overall jet energy scale, and flavor dependence. 
We consider all other uncertainties
to be fully correlated.  The combined measurement yields
$m_t=174.0\pm2.4\thinspace(\rm stat)\pm
1.4\thinspace(\rm syst)$~GeV.  This is consistent with measurements in
other channels, and is the most precise single $m_t$ measurement in
the dilepton channel to date. We have also improved
the precision by combining the \nuwt\ results 
with the results of Ref.~\cite{llmtopME}.
The statistical correlation of these
two measurements is approximately 60\%, calculated from 
pseudoexperiments.  Accounting for this
correlation, and 
correlations appropriate to each source of systematic
uncertainty, we obtain $m_t = 173.9 \pm 1.9
\thinspace(\rm stat) \pm 1.6\thinspace(\rm syst)$~GeV.

%
We thank the staffs at Fermilab and collaborating institutions,
and acknowledge support from the
DOE and NSF (USA);
CEA and CNRS/IN2P3 (France);
FASI, Rosatom and RFBR (Russia);
CNPq, FAPERJ, FAPESP and FUNDUNESP (Brazil);
DAE and DST (India);
Colciencias (Colombia);
CONACyT (Mexico);
NRF (Korea);
CONICET and UBACyT (Argentina);
FOM (The Netherlands);
STFC and the Royal Society (United Kingdom);
MSMT and GACR (Czech Republic);
BMBF and DFG (Germany);
SFI (Ireland);
The Swedish Research Council (Sweden);
and
CAS and CNSF (China).
%


\begin{thebibliography}{99}
\bibitem{pseudorapidity} 
  The pseudorapidity is defined as $\eta = - \ln[\tan(\theta /2)]$ where $\theta$ is the polar angle 
  relative to the beam axis, defined relative to the center of the detector.
\end{thebibliography}
\end{document}